\documentclass[prd,aps,showpacs,11pt,%showpacs,
nofootinbib,%
tightenlines,
notitlepage,
]{revtex4-1}

\usepackage{graphicx}
\usepackage{amsmath}
\usepackage{amssymb}
\usepackage{hyperref} 

\usepackage{color}
\usepackage{colortbl}
\newcommand{\be}{\begin{equation}}
\newcommand{\ee}{\end{equation}}
\newcommand{\bea}{\begin{eqnarray}}
\newcommand{\eea}{\end{eqnarray}}

\newcommand{\6}{\partial }

\newcommand{\MKK}{M_{\rm KK}}

\newcommand{\Tr}{{\rm tr}\,}

\newcommand{\AD}{$\overline{\text{D8}}$}

\newcommand{\psia}{\psi^A}
\newcommand{\zg}{z_>}
\newcommand{\zk}{z_<}
\newcommand{\zo}{z_0}

\begin{document}

\title{Axial vector transition form factors in holographic QCD
and their contribution to the 
anomalous magnetic moment of the muon
}

\author{Josef Leutgeb}
\author{Anton Rebhan}
\affiliation{Institut f\"ur Theoretische Physik, Technische Universit\"at Wien,
        Wiedner Hauptstrasse 8-10, A-1040 Vienna, Austria}

\date{\today}

\begin{abstract}
We evaluate axial vector transition form factors in holographic QCD models
that have been shown to reproduce well recent experimental and theoretical results for the pion
transition form factor. Comparing with L3 data on $f_1\to\gamma\gamma^*$
we find remarkable agreement regarding the shape of single-virtual form factors.
In the double-virtual case, the holographic results differ strongly
from a simple dipole form, 
and this has an important impact on the corresponding estimate of
the axial vector contribution to the anomalous magnetic moment of the muon $a_\mu$
through hadronic light-by-light scattering. 
We demonstrate that hard-wall models satisfy the 
Melnikov-Vainshtein short-distance constraint for the latter, if and only if the infinite
tower of axial vector states is included. The results for $a_\mu$, however, 
are strongly dominated by the first few resonances. Numerically, these
results turn out to be surprisingly large: (2.9 - 4.1)$\times 10^{-10}$ in the hard-wall models,
57-58\% of which are due to the longitudinal contribution, which is the one
responsible for the Melnikov-Vainshtein short-distance constraint.
Rescaling the holographic result to obtain an optimal fit of L3 data,
but then matching only 52\% of the asymptotic constraint,
the result is reduced to $2.2(5)\times 10^{-10}$, which is still
significantly larger than most previous phenomenological estimates 
of the axial vector exchange contribution.
\end{abstract}
%\pacs{11.25.Tq,13.25.Jx,14.40.Be,14.40.Rt}

\maketitle

\section{Introduction and summary}

Presently, there is a discrepancy between the measured and the predicted value of
the anomalous magnetic moment of the muon \cite{Jegerlehner:2017gek} of the order of \cite{Jegerlehner:2017lbd,Davier:2019can,Keshavarzi:2019abf}
$a_\mu^{\rm exp.}-a_\mu^{\rm theory} \simeq 26\times 10^{-10}$, above three standard deviations
with currently estimated errors.
In view of the upcoming new experiment at FERMILAB \cite{Abe:2019thb},
much effort is being put into reducing the theoretical uncertainty of the Standard Model prediction, which
is dominated by hadronic effects \cite{Prades:2009tw,Kurz:2014wya,Colangelo:2014qya,Jegerlehner:2017lbd,Davier:2019can,Keshavarzi:2019abf,Hoferichter:2018kwz,Danilkin:2019mhd}
(with interesting recent progress in lattice QCD \cite{Giusti:2019xct,Davies:2019efs,Gerardin:2019vio,Gerardin:2019rua,Blum:2019ugy,Borsanyi:2020mff}), while
QED \cite{Aoyama:2017uqe} and electroweak effects \cite{Czarnecki:2002nt,Gnendiger:2013pva} appear
under control (see \cite{Jegerlehner:2017gek,Danilkin:2019mhd} for more references). 

Although being smaller than the effects of hadronic vacuum polarization,
the hadronic light-by-light (HLBL) scattering contribution has %the largest 
a comparable uncertainty.
There the exchange of single pseudoscalar mesons $P=\pi^0,\eta,\eta'$ is the most important
contribution. For the latter recent advances have been made in particular using dispersion relations
\cite{Colangelo:2015ama,Hoferichter:2018kwz} and lattice QCD \cite{Gerardin:2019vio} 
to determine the all-important pseudoscalar transition form factors (TFF)
$P\to\gamma^*\gamma^*$, for which direct experimental information is available almost
exclusively in the single-virtual case. However, the HLBL contribution involves both a single-virtual
and a double-virtual TFF (in the external and the internal vertex, respectively).

In Ref.~\cite{Leutgeb:2019zpq}, we have recently revisited the predictions of chiral holographic QCD models 
\cite{Grigoryan:2007wn,Grigoryan:2008up,Grigoryan:2008cc,Hong:2009zw,Cappiello:2010uy}.
While these models are certainly only a crude approximation to real QCD (also in the chiral limit),
we found that the bottom-up holographic models introduced in \cite{Erlich:2005qh,DaRold:2005mxj,Hirn:2005nr}
agree remarkably well with new recent low-energy data \cite{Danilkin:2019mhd} for $\pi\to\gamma\gamma^*$
as well as with the results of the dispersive approach for double-virtual pion TFF \cite{Hoferichter:2018kwz},
leading to a result \cite{Leutgeb:2019zpq} for $a_\mu^{\pi^0}\simeq 5.9(2)\times 10^{-10}$ which is close to the new evaluations in 
\cite{Hoferichter:2018kwz,Danilkin:2019mhd,Gerardin:2019vio}.
The bottom-up holographic models with asymptotic anti-de Sitter geometry
can be matched to reproduce the leading-order (LO) perturbative QCD (pQCD)
short-distance constraint (SDC) of the vector current two-point function and then
even reproduce the exact form of the asymmetry function 
\be\label{fw}
f(w)=\frac1{w^2}-\frac{1-w^2}{2w^3}\ln\frac{1+w}{1-w},\quad w=(Q_1^2-Q_2^2)/(Q_1^2+Q_2^2),
\ee
in the LO pQCD
limit of the double-virtual TFF $F(Q_1^2,Q_2^2)$, where $Q_{1,2}^2$ are photon virtualities \cite{Grigoryan:2008up}.

An important SDC established by Melnikov and Vainshtein (MV) \cite{Melnikov:2003xd}
for the four-photon amplitude in the special limit $Q_1^2\sim Q_2^2 \gg Q_3^2 \to\infty$
is however missed by the pseudoscalar pole contribution to HLBL
(unless the single-virtual pion TFF at the external vertex is artificially eliminated and replaced
by its on-shell value, a procedure which was proposed in \cite{Melnikov:2003xd} as a simple model
to estimate the effect of incorporating the MV-SDC).

Encouraged by the success of bottom-up holographic models in treating the pion-pole contribution,
we consider here the axial vector contributions arising from the 5-dimensional Chern-Simons action,
which is responsible for the correct inclusion of the axial anomaly and which in the chiral
holographic models involves one pseudoscalar multiplet and an infinite tower of vector
and axial vector mesons. We verify that the latter are indeed responsible for satisfying the
MV-SDC, if these complete towers of (axial) vector mesons are included; any truncation
leads to a violation at infinite momenta. Nevertheless,
in the final result for $a_\mu$ only the first few multiplets of axial vector mesons
contribute significantly, with the lowest one yielding about 80\% of the complete result.

The TFF of the lightest isoscalar axial vector mesons that are predicted by the holographic
models can in fact be compared to experimental data from the L3 Collaboration \cite{Achard:2001uu,Achard:2007hm}.
Doing so, we find that the predicted $Q^2$ dependence of the single-virtual
TFF agrees perfectly with the data, when
the parameters of the holographic models are fixed to reproduce $f_\pi$ and $m_\rho$ (as was done
in our study of the pion TFF \cite{Leutgeb:2019zpq}).
With the latter, the hard-wall model of Ref.~\cite{Hirn:2005nr} (called HW2 below) reproduces the MV-SDC parametrically,
but numerically only at the level of 62\%, while the model of
Refs.~\cite{Erlich:2005qh,DaRold:2005mxj} (HW1), which has one more free parameter, can be made
to saturate it fully.
These models therefore provide a plausible extrapolation of the single-virtual TFF to the double-virtual
case needed for evaluating the axial vector contributions to $a_\mu$.
Using a simple dipole ansatz, Pauk and Vanderhaeghen (PV) \cite{Pauk:2014rta}
have extrapolated the experimental data for the single-virtual case to estimate $a_\mu^{f_1}$.
The holographic results turn out to have a very different asymptotic behavior and yield
much larger contributions.
The HW2 model, which reaches 62\% of the asymptotic MV-SDC
but agrees well with the low-energy normalization of the axial vector TFF extracted from experiment,
yields $a_\mu^\mathrm{AV}\approx 2.9 \times 10^{-10}$, while the HW1 model, which satisfies 100\% of the MV-SDC
but overestimates the low-energy normalization, gives approximately $4.1 \times 10^{-10}$.
Approximately 58\% and 57\% of these results (1.7 - $2.3\times 10^{-10}$)
arise from the longitudinal part of the axial vector meson propagator that
is responsible for the MV-SDC. 
Coincidentally, this is comparable to (albeit smaller than) the extra contribution obtained originally in 
the MV model \cite{Melnikov:2003xd}, $\Delta a_\mu^\mathrm{PS,MV}=
2.35\times 10^{-10}$,
where one structure function is artificially kept fixed to its on-shell value. However, 
when the MV model is updated to current input data \cite{Colangelo:2019lpu,*Colangelo:2019uex}, 
this increases to $\Delta a_\mu^\mathrm{PS,MV}=
3.8\times 10^{-10}$.
Above all, our holographic QCD study (together with our previous evaluation of the pseudoscalar pole contribution \cite{Leutgeb:2019zpq})
indicates that the simple MV model is not the correct way to implement the MV-SDC,
but that additional degrees of freedom are needed for that.

In Ref.~\cite{Colangelo:2019lpu,*Colangelo:2019uex} a different approach was recently taken to include
the MV-SDC by means of an infinite tower of pseudoscalar states (but vehemently criticized in \cite{Melnikov:2019xkq}). There the obtained
estimate 
for the effects of the MV-SDC was
smaller than our results, $1.3(6)\times 10^{-10}$. As discussed in \cite{Colangelo:2019lpu,*Colangelo:2019uex},
in the chiral large-$N_c$ limit only the lightest pseudoscalar states contribute, while an infinite
tower of axial vector mesons is present. The latter were not considered further in \cite{Colangelo:2019lpu,*Colangelo:2019uex}
on the grounds that they are poorly understood so far and that a good theoretical framework for treating them was missing.
The results obtained here demonstrate that they are naturally included in holographic QCD,
leading to a somewhat larger estimate of the effects of the MV-SDC than obtained in the model of
\cite{Colangelo:2019lpu,*Colangelo:2019uex}.

This paper is organized as follows. In the next section we briefly recapitulate the holographic hard-wall models
already discussed in our previous work on the pion TFF \cite{Leutgeb:2019zpq}, including for comparison
also the top-down model of Sakai and Sugimoto, which is found to compare well with low-energy results for the axial
vector TFF (in particular the experimental result for its normalization) while missing the SDC.
In Sect.\ \ref{sec:AVTFF} we first compare with the single-virtual results from the L3 Collaboration \cite{Achard:2001uu,Achard:2007hm}.
We then display the axial vector TFF also for the double-virtual case, highlighting its difference from the simple
model used in Ref.~\cite{Pauk:2014rta}, and work out its asymptotic behavior.
In Sect.\ \ref{sec:SDC} we show that the MV-SDC becomes satisfied when the complete infinite tower of axial
vector meson contributions is summed, while each individual contribution decays too fast to do so, and in
Sect.\ \ref{sec:amu} we finally evaluate the contributions to $a_\mu$.
Since in real QCD, away from the chiral large-$N_c$ limit, both excited pseudoscalar mesons and axial vector mesons
contribute, we also consider a data-driven adjustment of the holographic results, which are thus used as
a mere, albeit sophisticated phenomenological model for the axial vector TFF and the resulting contribution
for $a_\mu$, and which 
could be combined with models for excited pseudoscalar mesons along the lines of Ref.~\cite{Colangelo:2019lpu,*Colangelo:2019uex}.

\section{Holographic QCD models}

The AdS/CFT conjecture \cite{Maldacena:1997re} has led to many applications in strongly interacting non-Abelian gauge theories
in the limit of large color number $N_c$. 
In a top-down string-theoretic approach using type-IIA supergravity, Witten \cite{Witten:1998zw}
has shown that by a supersymmetry-breaking compactification one can construct a model of low-energy QCD
based on the near-horizon geometry of
D4 branes. Sakai and Sugimoto (SS) \cite{Sakai:2004cn,Sakai:2005yt} have extended this model by 
introducing $N_f$ probe D8 and anti-D8 branes localized in the extra dimension of the Witten model,
leading to a geometrical realization of chiral symmetry breaking of the $U(N_f)_\mathrm{L}\times  U(N_f)_\mathrm{R}$
symmetry of the unconnected branes, which in the confining geometry are forced to join in the bulk of the higher-dimensional
spacetime.

While the SS model is at best a model of low-energy QCD at large $N_f$, with no conformal symmetry emerging at high momentum scales,
simpler so-called bottom-up models have been constructed that break conformal symmetry by either a hard \cite{Erlich:2005qh,DaRold:2005mxj} 
or a soft wall \cite{Karch:2006pv}
in the bulk. The flavor gauge fields corresponding to chiral symmetry breaking are then introduced by hand
and are subjected to appropriate boundary conditions on these walls.

Thus both the top-down and the various bottom-up models eventually describe vector and axial
vector mesons through a $U(N_f)\times U(N_f)$
Yang-Mills action in a curved five-dimensional background [with or without a nontrivial dilaton ($\Phi$) background],
\be
S_{\rm YM} \propto \;\text{tr}\int d^4x \int_0^{z_0} dz\,e^{-\Phi(z)}\sqrt{-g}\, g^{PR}g^{QS}
\left(\mathcal{F}^\mathrm{L}_{PQ}\mathcal{F}^\mathrm{L}_{RS}
+\mathcal{F}^\mathrm{R}_{PQ}\mathcal{F}^\mathrm{R}_{RS}\right),
\ee
where $P,Q,R,S=0,\dots,3,z$ and $\mathcal{F}_{MN}=\partial_M \mathcal{B}_N-\partial_N \mathcal{B}_M-i[\mathcal{B}_M,\mathcal{B}_N]$.

In the SS model the D8 brane action also involves a Chern-Simons term, which leads to the correct Wess-Zumino-Witten term \cite{Sakai:2004cn,Sakai:2005yt}
\be\label{SCS}
S_{\rm CS}=\frac{N_c}{24\pi^2}\int\text{tr}\left(\mathcal{B}\mathcal{F}^2-\frac{i}2 \mathcal{B}^3\mathcal{F}
-\frac1{10}\mathcal{B}^5\right). 
\ee
In the bottom-up models, where $\mathcal{B}^\mathrm{L}$ and $\mathcal{B}^\mathrm{R}$ fields appear separately, 
the action (\ref{SCS}) is added by hand as $S_{\rm CS}^\mathrm{L}-S_{\rm CS}^\mathrm{R}$.
The electromagnetic gauge field can be introduced as a non-dynamical
background field through a nonzero boundary value for the vector gauge field
with generator equal to the electric charge matrix, which naturally leads to vector meson dominance (VMD) \cite{Sakai:2005yt}.

%%%%%%
As shown in Ref.~\cite{Son:2010vc}, Eq.~(\ref{SCS}) implies the correct leading SDC for the structure functions $w_{T,L}(Q^2)$ in
the %AVV 
vertex function of two vector and one axial-vector currents, $w_L(Q^2)=2N_c/Q^2$ (which is exact in the chiral limit), and $w_T(Q^2)=N_c/Q^2$, which
does not get perturbative corrections \cite{Vainshtein:2002nv}, 
but does receive nonperturbative contributions \cite{Czarnecki:2002nt,Knecht:2003xy} that are suppressed
by higher inverse powers of $Q^2$, which in holographic QCD depend
on the model \cite{Son:2010vc,Colangelo:2011xk}.

In the following we recapitulate the relevant formulae for the various models that we will use for deriving the axial vector TFF
and their contribution to $a_\mu$.
For more details see Ref.~\cite{Leutgeb:2019zpq} and
references therein.

\subsection{Sakai-Sugimoto model}

With a dimensionless coordinate $Z$ along the connected D8-\AD\ branes and holographic boundary at $Z=\pm\infty$,
the Yang-Mills part of the action of the SS model reads \cite{Sakai:2004cn,Sakai:2005yt}
\be\label{SD8F2}
S_{\rm YM}=-\kappa\,\Tr\int d^4x \int_{-\infty}^\infty dZ\left[
\frac12 (1+Z^2)^{-1/3}\eta^{\mu\rho}\eta^{\nu\sigma}\mathcal{F}_{\mu\nu}\mathcal{F}_{\rho\sigma}
+(1+Z^2)\MKK^2\eta^{\mu\nu}\mathcal{F}_{\mu Z}\mathcal{F}_{\nu Z}\right]
\ee
with $\kappa={\lambda N_c}/(216\pi^3)$ and $\lambda=g_{\rm YM}^2 N_c$.

An infinite tower of massive vector and axial vector mesons arises from even and odd eigenmodes
of $\mathcal{B}_\mu^{(n)}=\psi_n(Z) v^{(n)}_\mu(x)$ with eigenvalue equation
\be\label{psin}
-(1+Z^2)^{1/3}\6_Z\left[ (1+Z^2)\6_Z \psi_n \right]=\lambda_n \psi_n,\quad \psi_n(\pm\infty)=0.
\ee
The lowest mode $v_\mu^{(1)}$ is 
interpreted as the isotriplet $\rho$ meson (or the $\omega$ meson
for the U(1) generator) with mass $m_\rho^2=\lambda_1 \MKK^2$.
The numerical result $\lambda_1=0.669314\ldots$
fixes the Kaluza-Klein mass of the SS model to $\MKK=1.2223 m_\rho$.

The holographic pion mode function is associated 
with the derivative of the non-normalizable zero-mode of (\ref{psin}) of the axial vector sector,
$
\alpha^{\rm SS}(Z)=\frac{\pi}2 \arctan(Z).
$
Multiplied with a massless pseudoscalar field in Minkowski space, this appears as the field $\mathcal{B}_Z$ or, when
the radial gauge $\mathcal{B}_Z=0$ is used, in nontrivial boundary conditions on $\mathcal{B}_\mu$
\cite{Sakai:2004cn}.
The pion decay constant is given by $f_\pi^2=\lambda N_c\MKK^2/(54\pi^4)$ so that
choosing $f_\pi=92.4$ MeV corresponds to $\kappa=0.00745$ or $\lambda\approx 16.63$ for $N_c=3$.

A background electromagnetic field $A_\mu(x)$ can be included by setting $\psi(\pm\infty)=1$ for 
$\mathcal{B}_\mu=\mathcal{Q}A_\mu(x)\psi(Z)$ with $\mathcal{Q}=e\,{\rm diag}(\frac23,-\frac13,-\frac13)$. 
A real photon with $q^2=0$
corresponds to the trivial solution $\psi(Z)\equiv 1$, whereas a virtual photon with spacelike
momentum $Q^2>0$ is described by a solution where $\lambda_n\to -Q^2/\MKK^2$.
This defines the so-called bulk-to-boundary propagator $\mathcal{J}$, which is determined by
\be
(1+Z^2)^{1/3}\6_Z\left[ (1+Z^2)\6_Z \mathcal{J} \right]=\frac{Q^2}{\MKK^2}\mathcal{J},\quad \mathcal{J}(Q,Z=\pm\infty)=1.
\ee

At order $1/N_c$ the axial U(1)$_A$ is broken in the SS model, which thereby 
includes a Witten-Veneziano mechanism \cite{Witten:1979vv,Veneziano:1979ec} for giving mass to
the $\eta_0$ pseudoscalar according to \cite{Sakai:2004cn,Bartolini:2016dbk,Leutgeb:2019lqu} $m_{0}^2=N_f\lambda^2\MKK^2/(27\pi^2 N_c)$.
When explicit mass terms are added \cite{Aharony:2008an,Hashimoto:2008sr}, this indeed gives
the right ballpark to account for realistic pseudoscalar meson masses \cite{Brunner:2015oga}.

However, since the SS model is not asymptotically AdS, as it has a diverging dilaton in the UV,
it can serve as a holographic model of QCD only at small momenta.

\subsection{Hard-wall models}

In the hard-wall models of Refs.~\cite{Erlich:2005qh,DaRold:2005mxj,Hirn:2005nr}, the background geometry
is instead chosen as pure AdS with metric
\be
ds^2=z^{-2}(\eta_{\mu\nu}dx^\mu dx^\nu - dz^2)
\ee
(conformal boundary at $z=0$),
but with a cutoff at some finite value of the radial coordinate $z_0$.

The action for the flavor gauge fields reads
\be
S_{\rm YM} = -\frac{1}{4g_5^2} \int d^4x \int_0^{z_0} dz\,%e^{-\Phi(z)}
\sqrt{-g}\, g^{PR}g^{QS}
\,\text{tr}\left(\mathcal{F}^\mathrm{L}_{PQ}\mathcal{F}^\mathrm{L}_{RS}
+\mathcal{F}^\mathrm{R}_{PQ}\mathcal{F}^\mathrm{R}_{RS}\right),
\ee
where $P,Q,R,S=0,\dots,3,z$ and $\mathcal{F}_{MN}=\partial_M \mathcal{B}_N-\partial_N \mathcal{B}_M-i[\mathcal{B}_M,\mathcal{B}_N]$.

\subsubsection{Hard-wall model with bi-fundamental scalar (HW1)}

In Refs.~\cite{Erlich:2005qh,DaRold:2005mxj}, a bi-fundamental bulk scalar $X$
is introduced, with a five-dimensional mass term determined by the scaling dimension $\Delta=3$ of the chiral-symmetry
breaking order parameter $\bar q q$ of the boundary theory,
\be
S_X=\int d^4x \int_0^{z_0} dz\,\sqrt{-g}\,\text{tr}\left(
|DX|^2+3|X|^2 \right),
\ee
where $DX=\partial X-i \mathcal{B}^\mathrm{L} X+iX\mathcal{B}^\mathrm{R}$ and $X=U(x,z)v(z)/2$ with
$v(z)=m_q z+\sigma z^3$, where $m_q$ is the quark mass and $\sigma$ the quark condensate.

At a finite value $z_0$, a cutoff of AdS$_5$ space is
imposed with boundary conditions $\mathcal{F}^\mathrm{L,R}_{z\mu}=0$.

Vector mesons have holographic wave functions given by
\be\label{psinHW}
\partial_z\left[\frac1z \partial_z \psi_n(z)\right]+\frac1z M_n^2 \psi_n(z)=0
\ee
with boundary conditions $\psi_n(0)=\psi'_n(z_0)=0$, solved by $\psi_n(z)\propto zJ_1(M_n z)$ with $M_n$ determined by
the zeros of the Bessel function $J_0$, denoted by $\gamma_{0,n}$. 
Identifying $M_1=m_\rho=775$ MeV, we obtain
\be
z_0=\gamma_{0,1}/m_\rho=3.103 \, \text{GeV}^{-1}.
\ee

The vector bulk-to-boundary propagator is obtained by replacing $M_n^2\to -Q^2$ and choosing the boundary conditions
$\mathcal{J}(Q,0)=1$ and $\partial_z \mathcal{J}(Q,z_0)=0$, which gives
\be\label{HWVF}
\mathcal{J}(Q,z)=
Qz \left[ K_1(Qz)+\frac{K_0(Q z_0)}{I_0(Q z_0)}I_1(Q z) \right].
\ee

The coupling constant $g_5$ can be fixed by requiring that the vector current two-point function matches
the pQCD result \cite{Erlich:2005qh}
\be\label{PiVas}
\Pi_V(Q^2)=-\frac1{g_5^2 Q^2} \left( \frac1z \partial_z \mathcal{J}(Q,z) \right)\Big|_{z\to0}=-\frac{N_c}{24\pi^2}\ln Q^2,
\ee
leading to $g_5^2=12\pi^2/N_c$.

In the chiral limit, the holographic wave functions of the axial vector mesons are given by
\be\label{psianHW1}
\partial_z\left[\frac1z \partial_z \psia_n(z)\right]-g_5^2\sigma^2 z^3 \psia_n(z) +\frac1z (M^A_{n})^2 \psia_n(z)=0
\ee
with the same boundary conditions, $\psia_n(0)=\psia{}'_n(z_0)=0$.

The pion field appears as the longitudinal part of $\mathcal{B}^A_{M\parallel}=\partial_M\varphi$.
In the chiral limit, its holographic wave function can be given in closed form as 
$\phi(z)=\mathcal{J}^A(0,z)-1$, where $\mathcal{J}^A(Q,z)$ is the axial vector bulk-to-boundary
propagator with
\cite{Grigoryan:2007wn,Grigoryan:2008up}
\be
\mathcal{J}^A(0,z)=\Psi(z)=\Gamma({\textstyle{\frac23}})\left(\xi z^3/2\right)^{1/3}
\left[I_{-1/3}(\xi z^3)-\frac{I_{2/3}(\xi z_0^3)}{I_{-2/3}(\xi z_0^3)}I_{1/3}(\xi z^3) \right],
\ee
where $\xi=g_5\sigma/3$. % and $g_5^2=12\pi^2/N_c$.
The pion decay constant is determined by \cite{Erlich:2005qh}
\be\label{fpiPsi}
f_\pi^2=-\frac1{g_5^2}\left( \frac1z \partial_z \Psi(z) \right)\Big|_{z\to0}
\ee
yielding
\be\label{HW1sigma}
\frac{6\pi^2}{N_c}f_\pi^2=\frac{\Gamma(\frac23)}{\Gamma(\frac43)}\frac{I_{2/3}(\xi z_0^3)}{I_{-2/3}(\xi z_0^3)}(\xi/2)^{2/3}.
\ee
This fixes $\xi=(0.424 \,\text{GeV})^3$ for $f_\pi=92.4$ MeV.

\subsubsection{Hirn-Sanz model (HW2)}

The hard-wall model by Hirn and Sanz \cite{Hirn:2005nr} (called HW2 in \cite{Cappiello:2010uy,Leutgeb:2019zpq}) 
does not introduce a matrix-valued scalar field for the purpose of chiral symmetry breaking, but
imposes different boundary conditions for vector and axial vector mesons at $z_0$,
which correspond exactly to the relations that are obtained in the SS model at the point where D8 and anti-D8 branes meet.
Vector mesons are given by (\ref{psinHW}) with $\psi_n(0)=\psi'_n(z_0)=0$ as in the HW1 model, while axial vector mesons
satisfy the same eigenvalue equation but with $\psia_n(0)=\psia_n(z_0)=0$ and $\psia{}'_n(z_0)\not=0$.
This gives
\be
\psia_n(z) \propto z J_1(M^A_n z),\quad M^A_n=\gamma_{1,n}/z_0.
\ee
Since $z_0$ is already fixed by $m_\rho$, this leads to the prediction
$M^A_1/m_\rho=\gamma_{1,1}/\gamma_{0,1}=1.593\ldots$, which is very close to the experimental values $m_{a_1(1269)}/m_\rho \approx 1.587$
and $m_{f_1(1285)}/m_\omega \approx 1.638$. As shown in
Tables \ref{vmasses} and \ref{avmasses} the results for the lightest axial vector masses in the other models are also close, but
not as good,
%\footnote{\new{With $f_\pi$ kept at 92.4 MeV, it turns out that in the chiral HW1 model the lightest axial vector masses cannot be matched 
%exactly to experimental values even when the lightest vector mass is left unconstrained;
%$M^A_1$ then has a minimum of 1357 MeV for a value of $z_0$ that corresponds to $m_\rho\approx 890$ MeV.}} 
whereas excited (axial) vector masses turn out somewhat too high in all models.

\begin{table}%[t]
\bigskip
\begin{tabular}{l|c|c|c}
\toprule
 & $m_V$ & $m_{V^*}$ & $m_{V^{**}}$ \\
\colrule
SS  & \textit{775} &  1606.0 & 2379.3 \\
HW1,2  &  \textit{775} & 1778.9 & 2788.8 \\ 
HW2(UV-fit) & 987.2 & 2266.1 & 3552.6 \\
SW  &  \textit{775} & 1096.0 & 1342.3 \\
\colrule
$m_{\rho}$ (exp.) & 775.26(25) & 1465(25) & 1720(20) \\
$m_{\omega}$ (exp.) & 782.65(12) & 1425(25) & 1670(30) \\
$m_{\phi}$ (exp.) & 1019.461(19) & 1680(20) & \\
\botrule
\end{tabular}
\caption{
Holographic values of the masses of the three
lowest vector mesons in comparison with PDG data for the masses of
$\rho,\omega$ and $\phi$ mesons in MeV \cite{PDG18}.
All holographic models except HW2(UV-fit) are with $m_V=775$ as input;
HW2(UV-fit), which is used only
for the sake of comparison, is
the HW2 model with $g_5$ at $N_c=3$ matched to 
pQCD asymptotics, resulting in a smaller $z_0$
and thus much higher $m_V$.
}
\label{vmasses}
\end{table}

\begin{table}%[t]
\bigskip
\begin{tabular}{l|c|c|c}
\toprule
 & $m_A$ & $m_{A^*}$ & $m_{A^{**}}$ \\
\colrule
SS  & 1186.5 & 2019.8 & 2843.2 \\
HW1  &  1375.5 & 2154.2 & 2995.1 \\ 
HW2  &  1234.8 & 2260.9 & 3278.6 \\
HW2(UV-fit) & 1573.0 & 2880.1 & 4176.4 \\
SW1 & 1674.1 & 2669.2 & 3497.6 \\
\colrule
$m_{a_1}$ (exp.) & 1230(40) & 1655(16) & 1930($^{+30}_{-70}$) \\ %& 2270($^{+55}_{-40}$) \\
$m_{f_1}$ (exp.) & 1281.9(0.5) & 1670(30) & 1971(15) \\ %& 2310(60) \\
$m_{f'_1}$ (exp.) & 1426.3(0.9) &  & \\
\botrule
\end{tabular}
\caption{
Holographic values of the masses of the three
lowest axial vector mesons 
with fixed $m_\rho$
in comparison with PDG data for the masses of
$a_1$ and $f_1$ mesons in MeV \cite{PDG18}. 
(SW1 is the soft-wall version \cite{Kwee:2007dd} of HW1.)
}
\label{avmasses}
\end{table}

Similarly to the SS model, the pion field in the HW2 model is contained in Wilson lines running along the holographic direction, $U(x)=\xi_\mathrm{R}(x)\xi_\mathrm{L}(x)$
with $\xi_\mathrm{L,R}=P\exp(- i\int_0^{z_0}dz \mathcal{B}_z^\mathrm{L,R})$.
Its holographic wave function is determined by the axial vector bulk-to-boundary propagator at $Q^2=0$
\be
\mathcal{J}^A(0,z)=\Psi(z)=1-\frac{z^2}{z_0^2}
\ee
so that (\ref{fpiPsi}) yields
\be
g_5^2=\frac2{f_\pi^2 z_0^2}.
\ee

Having fixed $z_0$ by the $\rho$ meson mass, a realistic choice of $f_\pi=92.4$ MeV now leads to
a coupling $g_5\approx 4.932$ which is much smaller than the value $g_5=2\pi\sqrt{N_c/3}$ needed
at $N_c=3$ to match pQCD according to (\ref{PiVas}).

In Ref.~\cite{Leutgeb:2019zpq} we have seen that the HW2 model produces a pion TFF that agrees well
with existing experimental data when the model is matched to $m_\rho$ and $f_\pi$. 
The short-distance constraints for the pion TFF are then only satisfied at the level of 62\%.
Setting $g_5$ to match the short-distance constraint (\ref{PiVas}) would lead to
a $\rho$ meson mass of 987 MeV and also
strong discrepancies with the low-energy pion TFF data which are of crucial
importance to the HLBL contribution to $a_\mu$.  In the application to $a_\mu$ we shall
therefore keep the parameters obtained by matching the low-energy regime.
This is, however, still an important improvement over the SS model, since the short-distance
constraints are then satisfied at least qualitatively.
The HW2 model with UV-fitted
$g_5=2\pi$, $N_c=3$, and then $z_0=2.4359\,\text{GeV}^{-1}$ instead of $3.103\,\text{GeV}^{-1}$
is referred to as HW2(UV-fit) in Tables \ref{vmasses} and \ref{avmasses}.

Because of the extra parameter $\sigma$,
the HW1 model is able to incorporate both the desired low-energy parameters $m_\rho$ and $f_\pi$
as well as the full asymptotic pQCD limits. However, the latter are then presumably reached too quickly for $Q^2\gg m_\rho^2$
because pQCD corrections typically reduce leading order results by nonnegligible amounts at
most energy scales of interest. Taken together we might hope, however, that the HW1 and HW2 models
span a plausible range of predictions for real QCD.

In contrast to \cite{Cappiello:2010uy,Leutgeb:2019zpq}, where the pion TFF was studied, we do not include the soft-wall
model considered there. The original soft-wall model introduced in \cite{Karch:2006pv}
is very close to the HW1 model except that it introduces a nontrivial dilaton.
As shown in Table \ref{avmasses} this leads to a poor fit of
the mass of the lightest axial vector meson. Moreover the results of Ref.~\cite{Kwee:2007dd}
indicate that the HW1 model generally agrees better with pion data than the SW model.
In fact, in \cite{Cappiello:2010uy,Leutgeb:2019zpq}, following Ref.~\cite{Grigoryan:2008up}, 
a simplified version of the SW model
without bi-fundamental bulk scalar $X$ and an ad-hoc choice for the pion wave function
was used that does not cover the axial vector sector.

\section{Axial vector transition form factor}\label{sec:AVTFF}

\subsection{Holographic results}

The effective Lagrangian for the coupling of two photons with one axial vector meson arises from the Chern-Simons action (\ref{SCS})
after integrating over the holographic coordinate, which in the case of the SS model reads
\be
\mathcal{L}_{\mathcal{A}\gamma\gamma} %\subset
=-i\frac{N_c}{12\pi^2}\text{tr}\,\epsilon^{\mu\nu\rho\sigma}\int_{-\infty}^\infty dZ %\mathcal{B}_\mu \mathcal{B}'_\nu \partial_\rho\mathcal{B}_\sigma,
\left(a_\mu \mathcal{V}'_\nu \partial_\rho \mathcal{V}_\sigma
+\mathcal{V}_\mu a_\nu' \partial_\rho \mathcal{V}_\sigma+\mathcal{V}_\mu \mathcal{V}'_\nu \partial_\rho a_\sigma
\right),
\ee
where $a_\mu(x,Z)$ is the normalizable axial vector meson field and $\mathcal{V}_\mu(x,Z)$ is
a vector field whose boundary condition is the background photon field; a prime 
denotes differentiation with respect to $Z$.
With partial integrations this can be simplified to
\be\label{LeffAgg}
\mathcal{L}_{\mathcal{A}\gamma\gamma} %\subset
=-i\frac{N_c}{12\pi^2}\text{tr}\,\epsilon^{\mu\nu\rho\sigma} \left[\int_{-\infty}^\infty dZ
\left(-3 \mathcal{V}_\mu' a_\nu \partial_\rho \mathcal{V}_\sigma\right)
+\mathcal{V}_\mu a_\nu \partial_\rho \mathcal{V}_\sigma\Big|_{Z=-\infty}^\infty
\right].
\ee
In the HW models the integral over $Z$ becomes $2\int_0^{\zo} dz(\ldots)$.
The boundary term appearing in (\ref{LeffAgg}) vanishes in the SS model because $a_\mu(x,\pm\infty)=0$.
In the HW2 model, the corresponding boundary term also vanishes
because $a_\mu(x,0)=0=a_\mu(x,\zo)$, but not in the HW1 model, where $a_\mu(x,\zo)\not=0$.
Similarly to the case of the pion TFF in the HW1 model \cite{Grigoryan:2008up}, 
the resulting nonzero contribution at the infrared wall
needs to be subtracted from the Chern-Simons action. 
Otherwise one would obtain a nonzero amplitude for the decay of an axial vector meson
in two real photons (for which $\mathcal{V}_\mu$ is simply a constant with respect to the holographic coordinate),
and this would violate the Landau-Yang theorem \cite{Landau:1948kw,*Yang:1950rg}.
The latter is realized by the fact that $\mathcal{V}_\mu'$ in the integral in (\ref{LeffAgg}) vanishes 
when $Q^2\to0$.

We therefore write the
amplitude $\gamma^*(q_{(1)})\gamma^*(q_{(2)})\to \mathcal{A}^a$ for photon virtualities $Q_i^2=-q_{(i)}^2$ as
\be\label{calMa}
\mathcal{M}^a=i\frac{N_c}{4\pi^2}\mathrm{tr}(\mathcal{Q}^2 t^a)\,\epsilon_{(1)}^\mu \epsilon_{(2)}^\nu
\epsilon_{\mathcal{A}}^{*\rho} \epsilon_{\mu\nu\rho\sigma}
\left[q_{(2)}^\sigma Q_1^2 A(Q_1^2,Q_2^2)-q_{(1)}^\sigma Q_2^2 A(Q_2^2,Q_1^2)\right],
\ee
where
$\mathcal{Q}=e\,\text{diag}(\frac23,-\frac13,-\frac13)$ and the flavor matrices are
given by $t^a=\lambda^a/2$ and $t^0=\mathbf1/\sqrt6$.
The polarization vectors in (\ref{calMa}) are transverse to the respective four-momenta; writing $\mathcal{M}$
with photon polarization vectors removed, 
one should supply the corresponding projection operators. The form of (\ref{calMa}) is, however, such that no
factors $1/Q_{1,2}^2$ are left when doing so:
\bea\label{calMmunu}
\mathcal{M}_{\mu\nu}(q_{(1)},q_{(2)})&\propto&
\epsilon_{\mathcal{A}}^{*\rho} \epsilon_{\alpha\beta\rho\sigma}
\Bigl[
(q_{(1)}^2\delta^\alpha_\mu-q_{(1)}^\alpha q_{(1)\mu})q_{(2)}^\sigma\delta^\beta_\nu A(Q_1^2,Q_2^2)\nonumber\\
&& \qquad\qquad -(q_{(2)}^2\delta^\beta_\nu-q_{(2)}^\beta q_{(2)\nu})q_{(1)}^\sigma\delta^\alpha_\mu A(Q_2^2,Q_1^2)
\Bigr].
\eea
(See App.~\ref{AppMhel} for the resulting helicity amplitudes.)

In contrast to the model used in Ref.~\cite{Pauk:2014rta}, which assumes a similar form of $\mathcal{M}$,
the axial vector form factor $A$ following from (\ref{LeffAgg}) is not symmetric.\footnote{Even with asymmetric
form factor $A$, the form
(\ref{calMmunu}) is not the most general one permitted by gauge invariance. The latter admits another independent asymmetric
form factor, called $C$ in Ref.~\cite{Roig:2019reh}, which turns out to vanish for the Chern-Simons Lagrangian (\ref{LeffAgg}).}
In the HW models, it is given by
%In HW models:
\be %d_V'aV
\label{AHW}
A(Q_1^2,Q_2^2)= \frac2{Q_1^2} \int_0^{\zo} dz \left[ \frac{d}{dz} \mathcal{J}(Q_1,z) \right]
\mathcal{J}(Q_2,z) \psia(z) \Big/ \left[ g_5^{-2} \int_0^{\zo} \frac{dz}{z} (\psia)^2 \right]^{1/2},
\ee
whereas in the SS model we have
%In SS model:
\be %d_V'aV
A^\mathrm{SS}(Q_1^2,Q_2^2)= \frac1{Q_1^2} \int_{-\infty}^\infty dZ \left[ \frac{d}{dZ} \mathcal{J}^\mathrm{SS}(Q_1,Z) \right]
\mathcal{J}^\mathrm{SS}(Q_2,Z) \psia_\mathrm{SS}(Z) \Big/ \left[ \kappa \int_{-\infty}^\infty \frac{dZ}{Z} (\psia_\mathrm{SS})^2 \right]^{1/2}.
\ee
% Because $A(Q_1^2,Q_2^2)\not\equiv A(Q_2^2,Q_1^2)$ this could also be written as
% two form factors, a symmetrized one and an antisymmetrized one.

Because $\frac{d}{dz} \mathcal{J}(Q_1,z)$ vanishes like $Q_1^2$ in the limit $Q_1^2\to0$, these expressions
have a finite limit $A(0,0)$.
In the HW models, one obtains
\be
\lim_{Q\to0}\frac1{Q^2} \frac{d}{dz} \mathcal{J}(Q,z) = z \ln(z/\zo),
\ee
and in the SS model
\be
\lim_{Q\to0}\frac{\MKK^2}{Q^2} \frac{d}{dZ} \mathcal{J}^\mathrm{SS}(Q,Z) = \frac{Z}{1+Z^2}\,{}_2F_{1}(\frac13,\frac12;\frac32;-Z^2).
\ee

Using this, 
the result for $A(0,0)$ can be given in closed form for the HW2 model, reading for the lightest axial vector meson
\be 
A(0,0)=-4\sqrt{2} \frac{J_0(\gamma_{1,1})-1}{\gamma_{1,1}^3 J_0(\gamma_{1,1})}g_5 \zo^2 
=-0.3502 g_5 \zo^2 = -16.633\, \text{GeV}^{-2} \quad \text{(HW2)}
\ee
with $g_5=\sqrt{2}(f_\pi z_0)^{-1}\approx 4.932$. (With UV-fit at fixed $N_c$ and $f_\pi$, one would
obtain $-13.056\,\text{GeV}^{-2}$.) 

In the HW1 and SS models, the corresponding results have to be obtained numerically. %, and are given in Table ????.
For the former, we find
\be
A(0,0)=-0.3478 g_5 \zo^2 = -21.043\, \text{GeV}^{-2}
\quad \text{(HW1)}
\ee
with $g_5=\sqrt{12\pi^2/N_c}=2\pi$ and $z_0=3.103 \, \text{GeV}^{-1}$.
In the case of the SS model, the result is
\be
A(0,0)=-1.2379 \kappa^{-1/2} \MKK^{-2} = -15.926\, \text{GeV}^{-2}
\quad \text{(SS)}.
\ee

\subsection{Comparison with experimental data for $\gamma\gamma^*\to f_1$}

The above results for $A(0,0)$
can be compared with experimental data from the L3 Collaboration 
\cite{Achard:2001uu,Achard:2007hm} for the so-called equivalent two-photon decay width of the lightest $f_1$ mesons
with one quasi-real longitudinal photon of virtuality $Q_1^2$ and one real transverse photon \cite{Schuler:1997yw,Pascalutsa:2012pr},
\be
\tilde \Gamma_{\gamma\gamma}=\lim_{Q_1^2\to0} \Gamma(\mathcal A\to\gamma^*_L\gamma_T) M_A^2/(2Q_1^2).
\ee
This is related to the form factor $F^{(1)}_{\mathcal A\gamma^*\gamma^*}$ defined in Ref.~\cite{Pascalutsa:2012pr}
by
\be
\tilde \Gamma_{\gamma\gamma}= \frac{\pi\alpha^2 M_A}{12}[F^{(1)}_{\mathcal A\gamma^*\gamma^*}(0,0)]^2,
\ee
which in turn is related to $A(0,0)$ by
\be
M_A^{-2}  F^{(1)}_{\mathcal{A}\gamma^*\gamma^*}(0,0) = \frac{N_c}{4\pi^2}\mathrm{tr}(\mathcal{Q}^2 t^a) A(0,0),
\ee
where for $f_1$ mesons $t^a$ can be replaced by a mixture of singlet and octet generators.
Similarly to $\omega$ and $\phi$, the physical $f_1$ and $f_1'$ states are expected to be close to an ideal
mixing scenario, where $f_1$ is predominantly $\bar u u+\bar d d$, while $f_1'$ is mainly $\bar s s$.
The experimental results $\tilde \Gamma_{\gamma\gamma}=3.5(8)$ keV for $f_1(1285)$  \cite{Achard:2001uu} 
and 3.2(9) keV for $f_1(1420)$ \cite{Achard:2007hm}, which are fairly close numerically, indicate however
a certain deviation from ideal mixing; otherwise the radiative decay of $f_1'$ would be more strongly suppressed compared to $f_1$.
The mixing angle for the $f_1$--$f_1'$ system is usually defined as
\be\label{f1mixing}
|f_1(1285)\rangle=\cos \phi_f|\bar n n\rangle-\sin \phi_f|\bar s s\rangle
\ee
with $|\bar n n\rangle = (|\bar u u\rangle+|\bar d d\rangle)/\sqrt2$.
Assuming a universal value of $A(0,0)$, the experimental results for $\tilde \Gamma_{\gamma\gamma}$ imply $\phi_f\approx 20.4^\circ$,
which is close to the recent LHCb result \cite{Aaij:2013rja} of $\phi_f=\pm(24\pm3)^\circ$
and other results pointing to a range of +(20\ldots30)$^\circ$ \cite{Dudek:2011tt,Cheng:2011pb}.
Translated to $A(0,0)$, the experimental results \cite{Achard:2001uu,Achard:2007hm} thus imply a value of
\be
|A(0,0)|^\mathrm{exp.} \simeq 15(2) \text{GeV}^{-2}.
\ee
The above holographic results for the SS and HW2 models agree remarkably well with this, while the HW1 model appears to overestimate
the radiative decay amplitudes of the lightest $f_1$ mesons by 40\% (corresponding to a factor 2 for $\tilde\Gamma_{\gamma\gamma}$).

\begin{figure}
\includegraphics[width=0.65\textwidth]{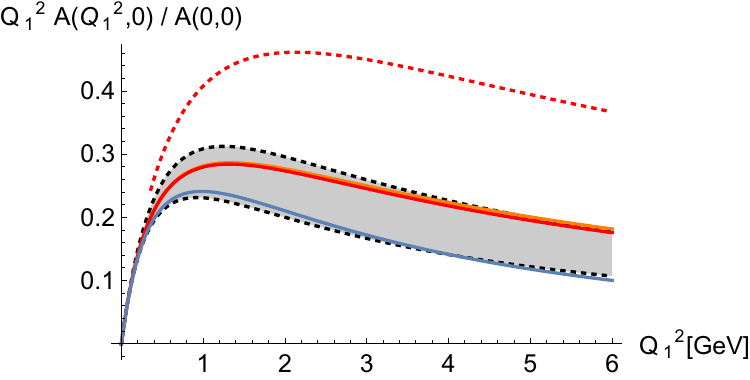}
\caption{Single-virtual axial vector TFF from holographic models 
(SS: blue, HW1: orange, HW2: red) compared with
dipole fit of L3 data for $f_1(1285)$ (grey band). The parameters
of all models are fixed by matching $f_\pi$ and $m_\rho$.
The results for HW1 and HW2 almost coincide, with HW2 at most a line
thickness above HW1. When the mass scale $\zo^{-1}$
is not fixed by $m_\rho$ but instead matched to the pQCD with $N_c=3$, HW2(UV-fit) instead gives
the significantly larger result denoted by the red dotted line.}
\label{fig:f1ffsingle}
\end{figure}

\begin{figure}
\includegraphics[width=0.6\textwidth]{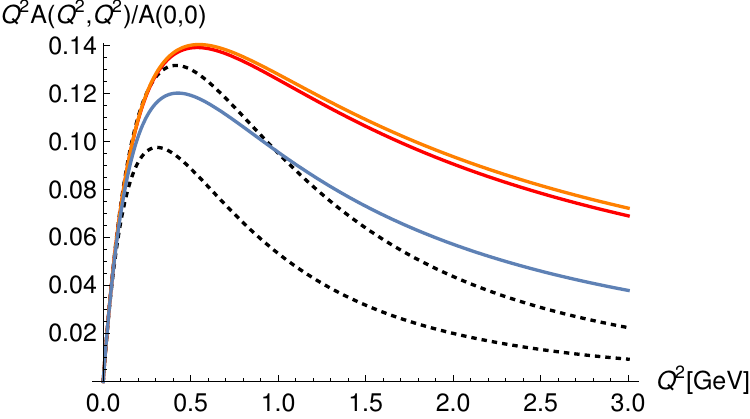}
\caption{Double-virtual axial vector TFF for $Q_1^2=Q_2^2=Q^2$ from holographic models 
(SS: blue, HW1: orange, HW2: red). The black dashed lines denote the extrapolation
of L3 data with a dipole model for each virtuality as used in the calculation of $a_\mu^{f_1}$ in Ref.~\cite{Pauk:2014rta}.}
\label{fig:f1ffsym}
\end{figure}

There exist also data on the $Q^2$ dependence of $f_1\to \gamma\gamma^*$.
In the analysis of the L3 data \cite{Achard:2001uu,Achard:2007hm} the single-virtual TFF of the axial vector mesons
has been modeled by a dipole ansatz corresponding to 
\be
\frac{A(Q_1^2,0)}{A(0,0)}=\frac1{(1+Q_1^2/\Lambda_D^2)^2}
\ee
with $\Lambda_D=1040\pm78$ MeV and $926\pm78$ MeV for $f_1(1285)$ and $f_1(1420)$, respectively.

As will be discussed below, an asymptotic behavior $\sim Q_1^{-4}$ is indeed also implied by the holographic HW results,
which however have a more complicated form at moderate values of $Q_1$.
In Fig.\ \ref{fig:f1ffsingle} we compare the experimental fit for $f_1(1285)$
to the three holographic results, displaying a remarkable agreement with all of them
when their parameters are fixed to match the low-energy input parameters $f_\pi$ and $m_\rho$.\footnote{The results
for $f_1(1420)$ are lower than those $f_1(1285)$ with some overlap. However, since we consider
only chiral models in this paper, a comparison with the result for $f_1(1285)$ is more relevant, as the latter is
dominantly $\bar n n$.} (This agreement is however spoiled if the HW2 model is forced to exactly match
pQCD asymptotically, as also happens in the case of the pion TFF.)

In Fig.~\ref{fig:f1ffsym} we display the holographic results also for the double virtual TFF
with $Q_1^2=Q_2^2=Q^2$. In the calculation of the $f_1$ axial vector meson contributions to $a_\mu$ 
by Pauk and Vanderhaeghen \cite{Pauk:2014rta}, the experimental fit has been extrapolated to
\be\label{PVmodel}
\frac{A^{\mathrm{PV}}(Q_1^2,Q_2^2)}{A(0,0)}=\frac1{(1+Q_1^2/\Lambda_D^2)^2(1+Q_2^2/\Lambda_D^2)^2}.
\ee
This is represented by the dashed lines in Fig.~\ref{fig:f1ffsym}, which deviate rather strongly
from the holographic results. Indeed, the asymptotic behavior of the latter has the same
power-law in the single and the double virtual cases, as we discuss in the following, while (\ref{PVmodel})
decays like $Q^{-8}$ in the double virtual case.

\subsection{Asymptotic behavior}

Inspecting the asymptotic behavior in more detail, we introduce the definitions
\bea\label{eq:w}
&& Q_{1,2}^2=r^2_{1,2}Q^2 \equiv (1\pm w)Q^2,\nonumber\\ 
&& Q^2=\frac12 (Q_1^2+Q_2^2),\quad
w=(Q_1^2-Q_2^2)/(Q_1^2+Q_2^2).
\eea

In both HW models we obtain
\bea
A(Q_1^2,Q_2^2) &\to& 
\frac{a A(0,0)}{(\zo Q)^4} r_1^{{-1}} r_2 \int_0^\infty d\xi\,
\xi^3 
\left[K_1(r_1\xi)+r_1\xi K_1'(r_1\xi) \right]
K_1(r_2\xi)
\nonumber\\
&& = \frac{a {A(0,0)}}{(\zo Q)^4}
\frac1{w^4}\left[
w(3-2w)+\frac12 (w+3)(1-w)\ln\frac{1-w}{1+w}
\right]
\eea
with a dimensionless constant $a$ which differs between HW1 and HW2.
Higher axial vector modes have the same form, but with different $a$.
(The SW model has the same asymptotic $Q$ and $w$ dependence.)\footnote{As we learned through private communication from
Martin Hoferichter and now published in  \cite{Hoferichter:2020lap}, 
the same asymmetry function is obtained when the axial vector TFF is calculated
in pQCD with the Brodsky-Lepage formalism.}

\begin{figure}
\includegraphics[width=0.5\textwidth]{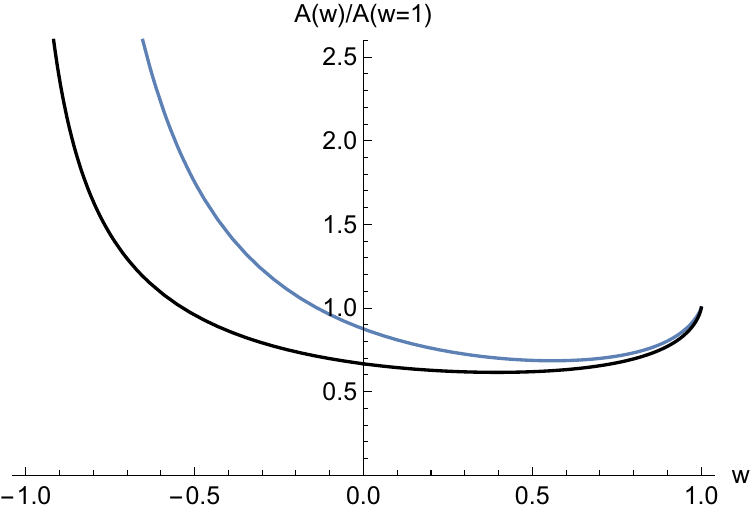}
\caption{Dependence of $A(Q_1^2,Q_2^2)$ on the asymmetry parameter $w$ (defined in (\ref{eq:w})) in the asymptotically AdS bottom-up models (black line)
and in the SS model (blue line). In the single-virtual limit, only $w=+1$ appears in scattering amplitudes.}
\label{fig:Aw}
\end{figure}

The $w$-dependence (displayed in Fig.\ \ref{fig:Aw}) is asymmetric with a minimum at $w \approx 0.395$
and a logarithmic singularity at $w=-1$,
corresponding to a behavior $A\sim \ln(Q\zo)/Q^4$ for $A(0,Q\to\infty)$.

The SS model, which cannot be matched to pQCD at high momentum scales, 
decays faster for $Q\to\infty$, with a qualitatively different
$w$-dependence. For completeness, the 
asymptotic behavior of the form factor $A$ in the SS model is given by
\bea
A^\mathrm{SS}(Q_1^2,Q_2^2) &\to& \frac{a^\mathrm{SS} A^\mathrm{SS}(0,0)\MKK^5}{Q^5}
r_1^{-1}\int_0^\infty d\xi\,
\xi^4 (1+3 r_2)e^{-3(r_1\xi+r_2\xi)} \nonumber\\
&&=\frac{a^\mathrm{SS} A^\mathrm{SS}(0,0)\MKK^5}{Q^5}\frac{8}{81}
\frac{6\sqrt{1-w}+\sqrt{1+w}}{\sqrt{1+w}(\sqrt{1-w}+\sqrt{1+w})^6},
\eea
and the form of this $w$ dependence is also displayed in Fig.\ \ref{fig:Aw}.
This has its minimum at $w\approx 0.560$ and a power-law singularity at $w=-1$,
corresponding to a behavior $A^\mathrm{SS}\sim 1/Q^4$ for $A^\mathrm{SS}(0,Q\to\infty)$,
so that in this particular limit the discrepancy with the bottom-up models is reduced
to a merely logarithmic one. Note, however, that the single virtual case $a_1/f_1\to\gamma\gamma^*$
corresponds to the other limit of $w=+1$.

\section{Axial vector contribution to the four-photon amplitude and longitudinal short-distance constraints}\label{sec:SDC}

In the Bardeen-Tung-Tarrach basis of the HLBL four-point function \cite{Colangelo:2015ama}, the short-distance constraint of Melnikov and
Vainshtein \cite{Melnikov:2003xd} for $N_f=3$ reads \cite{Colangelo:2019lpu,*Colangelo:2019uex}
\be\label{MVSDC}
\lim_{Q_3\to\infty} \lim_{Q\to\infty} Q^2 Q_3^2 \bar\Pi_1(Q,Q,Q_3)=-\frac{2}{3\pi^2}.
\ee

We shall now show that this constraint is satisfied in the HW models with $g_5^2=12\pi^2/N_c$,
which also ensures the correct short-distance limits of single and double virtual pion TFF.
In the HW1 model this can be achieved while fitting $m_\rho$ to its experimental value,
whereas in the HW2 model the latter together with $N_c=3$ needs a smaller value of
$g_5^2=2/(f_\pi z_0)^2$ so that the SDCs on the pion TFF are satisfied only at the level of 62\%  \cite{Leutgeb:2019zpq}.

However, the %MV-SDC 
constraint (\ref{MVSDC}) is only satisfied if the infinite tower of axial vector mesons
is taken into account; it is missed completely when only pions and a finite number of axial vector mesons are
included.

The axial vector contribution to $\bar\Pi_1(Q,Q,Q_3)$ comes from the longitudinal part of the axial
vector propagator $q^\mu_{(3)} q^\nu_{(3)}/(M^A_n Q_3)^2$ and has the form
\be\label{barPisum}
\bar\Pi_1=-\frac{g_5^2}{2\pi^4}
\sum_{n=1}^\infty 
\int_0^{\zo} dz \left[ \frac{d}{dz} \mathcal{J}(Q,z) \right]
\mathcal{J}(Q,z) \psia_n(z) \frac1{(M^A_n Q_3)^2}  \int_0^{\zo} dz' \left[ \frac{d}{dz'} \mathcal{J}(Q_3,z') \right]
\psia_n(z'),
\ee
where we have used that 
$
\sum_{a=0,3,8} \left( {N_c} \mathrm{tr}(\mathcal{Q}^2 t^a/e^2) \right)^2=1.
$
The results of the above section show that at $Q\to\infty$ the first integral appearing therein behaves
as $1/Q^2$ for $Q\to\infty$, and the second integral, which is a single-virtual
form factor, provides a factor of $1/Q_3^2$ multiplying a $1/Q_3^2$ from the propagator.
Thus each summand has a vanishing contribution to (\ref{MVSDC}).

However, the infinite sum behaves differently. This can be demonstrated in closed form in the HW2 model,
where both $\mathcal J$ and $\psia$ are given by Bessel functions.

%\subsection{Hirn-Sanz model}

With %incompletely normalized 
radial wave functions %/orthonormal w.r.t $\int_0^zo dz z^{-1}(\psia(z))^2_n=1$,
\be
\psia_n(z)=\sqrt2 z \zo^{-1} J_1(\gamma_{1,n}z/\zo)/|J_0(\gamma_{1,n})|
\ee
%(for convenience here 
normalized such that $\int_0^{\zo} dz z^{-1}(\psia(z))^2_n=1$
one has
\be
\sum_{n=1}^\infty \psia_n(z) \psia_n(z')=z\delta(z-z').
\ee

The sum $\sum_{n=1}^\infty \psia_n(z) \psia_n(z')/(M^A_n)^2$
is a special case of the axial vector bulk-to-bulk propagator.
In a mixed (Euclidean) 4-momentum and radial-coordinate representation the latter is given
by
\be
\sum_{n=1}^\infty \frac{\psia_n(z) \psia_n(z')}{Q^2+(M^A_n)^2}
=G^A(Q;z,z')
\ee
with
\be
G^A(Q;z,z')=z z' \left[K_1\left(Q\zg\right) I_1\left(Q\zo\right)-I_1\left(Q\zg\right) K_1\left(Q\zo\right)\right]
I_1\left(Q\zk\right)/I_1\left(Q\zo\right),
\ee
where $\zk=\mathrm{min}(z,z')$ and $\zg=\mathrm{max}(z,z')$.

\begin{figure}
\includegraphics[width=0.75\textwidth]{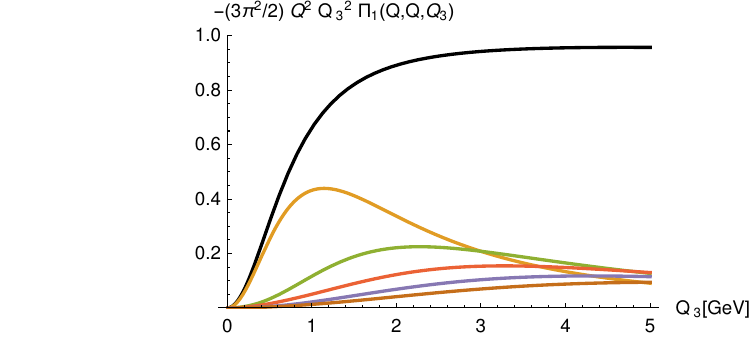}\qquad\qquad\qquad\qquad
\caption{Axial-vector contribution to
$Q_3^2 Q^2 \bar\Pi_1(Q,Q,Q_3)$ as a function of $Q_3$ at $Q=50$ GeV
in the HW2 model %%%%%%%%%%%%%%%%%%%%%%%%%%%%%%%%%% <---
normalized to the asymptotic value (\ref{MVSDC}) (with prefactor $g_5^2/(2\pi)^2$ set to one). 
The black line corresponds to the infinite sum over the tower of axial vector mesons, and the other lines
give the contributions of the 1st to 5th lightest axial vector mesons.
}
\label{fig:MV}
\end{figure}

The limiting case of $Q=0$ needed for the longitudinal contribution (\ref{barPisum}) is given by
the simple expression
\be\label{zGA}
G^A(0;z,z')=\frac{\zk^2(\zo^2-\zg^2)}{2\zo^2}.
\ee

Setting $Qz=\xi$, $Q_3 z'=\xi'$, the leading term in $G^A(0;z,z')$
at large momenta becomes $\zk^2/2=\text{min}(\xi^2/Q^2,\xi'^2/Q_3^2)/2$ so
that for $Q^2 \gg Q_3^2$ we obtain
\bea
&&-\frac{g_5^2}{2\pi^4}\frac1{2Q_3^2} \int_0^\infty d\xi \int_0^\infty d\xi'
\xi K_1(\xi)\frac{d}{d\xi}[\xi K_1(\xi)]\frac{d}{d\xi'}[\xi' K_1(\xi')]\xi^2/Q^2\nonumber\\ %%%%%% <---- ' on K_1(\xi') entfernt
&&={+}\frac{g_5^2}{2\pi^4}\frac1{2Q_3^2} \int_0^\infty d\xi  %%%%%%%% <---- sign corrected
\xi K_1(\xi)\frac{d}{d\xi}[\xi K_1(\xi)]\xi^2/Q^2=-\frac{2}{\pi^2}\frac1{2Q_3^2}\frac2{3Q^2}
\eea
for $g_5^2=4\pi^2$ at $N_c=3$, exactly reproducing the short-distance constraint (\ref{MVSDC}).
Notice that in this limit the single-virtual form factor effectively gets replaced by
$A(0,0)$, something that was done by hand in the model of Ref.~{\cite{Melnikov:2003xd} to
account for the short-distance constraint (albeit in the pseudoscalar sector, whereas
here this takes place exclusively in the case of axial vector mesons).

In Fig.~\ref{fig:MV}, the axial vector contribution to
$Q_3^2 Q^2 \bar\Pi_1(Q,Q,Q_3)$ is plotted for the HW2 model %(with prefactor $g_5^2/(2\pi)^2=1$) 
as a function of $Q_3$ at $Q=50$ GeV for $Q_3$ up to 5 GeV so that the
kinematic regime $Q^2\gg Q_3^2\gg m_\rho^2$ of the SDC (\ref{MVSDC})
is probed. The full result involving the infinite sum over the tower of axial vector
mesons is given by the black line, which is seen to approach the correct limit,
while each individual contribution decays for $Q_3\to\infty$.

In the chiral limit, the longitudinal SDC (\ref{MVSDC}) is in fact stronger and holds for all values of $Q_3$.
Including the pseudoscalar exchange contribution of the HW2 model \cite{Leutgeb:2019zpq} (but with vanishing pseudoscalar mass)
one can readily show that 
\be
\lim_{Q\to\infty} Q^2 \bar\Pi_1(Q,Q,Q_3)=-\frac{g_5^2}{(2\pi)^2}\frac{2}{3\pi^2 Q_3^2}
\ee
by partially integrating (\ref{barPisum}) and using that
\be\label{dzdzpGA}
\partial_{z'}\partial_z G^A(0;z,z')=-z \delta(z-z')-2zz'/z_0=
-z \delta(z-z')-\Psi'(z)\Psi'(z')/(2z_0^2),
\ee
thus verifying that the axial anomaly is correctly implemented.
We have checked numerically that this also holds true for the HW1 model, where $\mathcal J(Q,z)$ is unchanged but
the bulk-to-bulk propagator at zero momentum (\ref{zGA}) is replaced by\footnote{However, 
for $Q\not=0$, $G^A(0;z,z')$ can no longer be given in closed form but has to be constructed numerically.}
\be\label{zGAHW1}
G^A(0;z,z')=%\frac{\zk^2(\zo^2-\zg^2)}{2\zo^2}.
\frac{\pi}{3\sqrt3} zz' I_{1/3}(\xi \zk^3)
 \left[I_{-1/3}(\xi \zg^3)-\frac{I_{2/3}(\xi z_0^3)}{I_{-2/3}(\xi z_0^3)}I_{1/3}(\xi \zg^3) \right],
\ee
which does not obey a relation analogous to (\ref{dzdzpGA}).

\section{Axial vector contribution to $a_\mu$}\label{sec:amu}

Using the method of Gegenbauer polynomials in Ref.~\cite{Jegerlehner:2009ry}, %formalism of Ref.~\cite{}, 
the axial vector contribution to the four-photon amplitude
leads to an integral representation of the anomalous magnetic moment of the form (for details see App.~\ref{AppIntegrals})
\be\label{amuAVint}
a_\mu^\mathrm{AV}=\int_0^\infty dQ_1 \int_0^\infty dQ_2 \int_{-1}^1 d\tau \,\rho_a(Q_1,Q_2,\tau).
\ee
We have checked our master formula also using the formalism of
Refs.~\cite{Colangelo:2015ama,Colangelo:2017fiz}.

\begin{figure}
\includegraphics[width=0.65\textwidth]{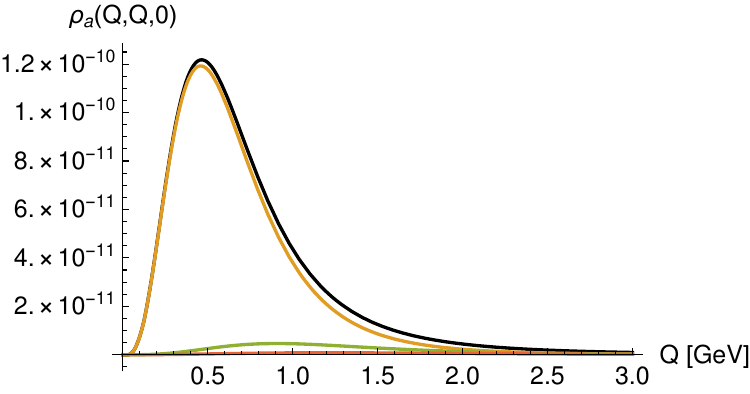}
\caption{The integrand $\rho_a(Q_1,Q_2,\tau)$ in (\ref{amuAVint}) in units of GeV$^{-2}$
for $Q_1=Q_2$ and $\tau=0$ (implying $Q_3=\sqrt{2}Q$) in case of the HW2 model. The black line is
the result from the infinite sum over the tower of axial vector mesons, the other lines
give the contributions of the 1st to 3rd lightest axial vector meson multiplets.
}
\label{fig:rhoaAV}
\end{figure}

In Fig.~\ref{fig:rhoaAV} we compare the integrand obtained with the full tower of axial vector mesons
in the HW2 model to the contribution of the first three multiplets at $Q_1=Q_2$ and $\tau=0$ (implying $Q_3=\sqrt{2}Q$).
This shows that the higher modes, which are essential for satisfying the MV-SDC, contribute only weakly
to $a_\mu$. 
In the integrated result for $a_\mu^\mathrm{AV}$, including only the lightest multiplets of 
axial vectors %up to and including mode number $j$, the HW2 model yields
gives about 80\% of the full result; after that each inclusion of one more multiplet roughly halves
the distance to the full result, see Table \ref{tab:amuj}.
A similar pattern holds for the HW1 model, where we do not have a closed form representation
of the infinite tower so that we had to resort to additional numerical estimates.\footnote{With (\ref{zGAHW1}) the longitudinal part
of the full integrand is given in closed form and therefore can be evaluated directly. The transverse contribution was estimated
by evaluation of the lowest seven modes and monitoring the ratio of transverse and longitudinal contributions, from which
it was concluded that the accurately determined longitudinal contribution amounts to 57\% of the full result.} 
There the lightest multiplet accounts for about 77\%.
By contrast, the SS model, which misses the MV-SDC completely and which we
will therefore discard in the following, has much smaller
contributions from the higher axial vector multiplets.

The numerical results for $a_\mu^\mathrm{AV}$ in the HW1 and HW2 models, whose parameters have been
fixed to reproduce $f_\pi$ and $m_\rho$, turn out to be surprisingly large, amounting to
roughly one half of the $\pi^0$ pole contribution obtained in our previous work \cite{Leutgeb:2019zpq}.
Approximately 58\% of the result $a_\mu^\mathrm{AV}$ is due to the contribution of the longitudinal part of
the axial vector meson propagator, i.e., (1.7 - 2.4)$\times 10^{-10}$ for the range spanned by HW2 and HW1.
This is smaller than the large increase $\Delta a_\mu^\mathrm{PS}|_\mathrm{MV}$
obtained in the simple model suggested in Ref.~\cite{Melnikov:2003xd} to satisfy the MV-SDC by
replacing the external form factor by a constant (which gave $2.35\times 10^{-10}$ in \cite{Melnikov:2003xd},
but would be $3.8\times 10^{-10}$ with current input data \cite{Colangelo:2019lpu,*Colangelo:2019uex}),
but somewhat higher than the recent estimate of \cite{Colangelo:2019lpu,*Colangelo:2019uex}
at $\Delta a_\mu^\mathrm{LSDC}=1.3(6)\times 10^{-10}$. %\footnote{
Reference \cite{Melnikov:2003xd} also applied this modification to the transverse
part of the axial vector meson contribution, leading to an enhancement by a factor of 2.7
of what their ansatz for the axial vector TFF would otherwise have given, resulting in the estimate $2.2\times 10^{-10}$
for the lowest axial vector meson multiplet. Our result for the transverse part of the latter amounts instead to
(1.2 - 1.7)$\times 10^{-10}$.

Even when only the lowest multiplet of axial vector mesons is included, we find a substantially
larger contribution than is typically estimated in the more recent literature \cite{Jegerlehner:2017gek,Pauk:2014rta}, where
longitudinal and transverse contributions from axial vector mesons are usually not separated.
Replacing the holographic form factors by the one used in Ref.~\cite{Pauk:2014rta}, Eq.~(\ref{PVmodel}),
we in fact reproduce the result $0.5\times 10^{-10}$ given therein as central value 
for $a_\mu[f_1(1285)]$. The main discrepancy arises
from the different asymptotic behavior of the form factor, which in Ref.~\cite{Pauk:2014rta}
is $Q^{-8}$ in the double virtual case. Another difference is the larger value of $|A(0,0)|$ obtained
in the holographic models, which appear to overestimate somewhat the equivalent two-photon decay width
$\tilde\Gamma_{\gamma\gamma}$. 
(In the case of the HW2 model, however, the holographic prediction of 4.2 eV
for $\tilde\Gamma_{\gamma\gamma}$ is completely within the experimental error ($3.5\pm0.8$ eV)
when the HW2 model is fitted to correct IR values rather than the pQCD asymptotics.)

If one uses only the holographic results for the normalized form factor $A(Q_1^2,Q_2^2)/A(0,0)$, where
the HW1 and HW2 models obtain very similar results, and adjusts the normalization $A(0,0)$ in order
to match the 
experimental result for $\tilde\Gamma_{\gamma\gamma}$, 
we find  % with mixing angle $\phi_f$ given below (\ref{f1mixing}) % $\phi_f=20.435^\circ$,
$a_\mu^{f_1}=0.88(20)\times 10^{-10}$
for $f_1(1285)$, in place of the result $0.5(2)\times 10^{-10}$ given in \cite{Pauk:2014rta}.
Extended to the complete $a_1,f_1,f_1'$ multiplet, 
this leads to %axial vector mass as given by the models
$a_\mu^\mathrm{AV1}=1.74(40)\times 10^{-10}$,
which can be viewed as a data-driven result, where holographic QCD is used as an interpolator
from single-virtual data to the double-virtual domain.\footnote{For future tests of the latter, we have included
in Appendix \ref{AppMhel} the expressions for the helicity amplitudes for photoproduction of axial
vector mesons.}
% could be further extended by adjusting to measured axial vector masses

The full axial vector exchange contribution will also involve higher multiplets, for which at present
no data are available.
Assuming that the lowest multiplet again accounts for 80\% of the total contribution, our
estimate for the latter is
$
a_\mu^\mathrm{AV}=2.2(5)\times 10^{-10}.
$

This downscaling of our holographic results to match experimental data implies that the MV-SDC is
satisfied to a lesser degree (unless it is applied only to a finite number of axial vector multiplets). 
The HW2 model with IR-fixed parameters, which reaches only 62\% of the MV-SDC,
needs only a moderate change, while the HW1 model, which saturates the MV-SDC, requires a much stronger one.
After such an overall downscaling, both HW models have almost exactly the same level of 52\% to which the MV-SDC is met.
This could perhaps be interpreted as an indication that in models that are closer to
real QCD other contributions may be of comparable importance, notably from the excited
pseudoscalar mesons \cite{Colangelo:2019lpu,*Colangelo:2019uex}, which in the chiral large-$N_c$ limit of our holographic models do not 
arise.

\begin{table}%[t]
\bigskip
\begin{tabular}{l|c|c|c|c|c|c}
\toprule
 & $j=1$ & $j\le2$ & $j\le3$ & $j\le4$ & $j\le5$ & $a_\mu^\mathrm{AV}$ \\
\colrule
HW1 & 3.14 & 3.62 & 3.79 & 3.91 & 3.96 & 4.06 $\times 10^{-10}$ \\
HW2 & 2.30 & 2.62 & 2.74 & 2.79 & 2.82 & 2.87 $\times 10^{-10}$ \\
HW2(UV-fit)& 2.37 & 2.69 & 2.81 & 2.86 &  2.89 & 2.94 $\times 10^{-10}$ \\
SS & 1.38 & 1.45 & 1.47 & 1.48 & 1.48 & 1.48 $ \times 10^{-10}$ \\
\botrule
\end{tabular}
\caption{The contribution of the infinite tower of axial vector mesons to $a_\mu^\mathrm{AV}$,
calculating in the HW2 model with the analytic expression for the bulk-to-bulk propagator but
estimated numerically for the HW1 model (and for comparison also for the SS model, which misses
the MV-SDC qualitatively). The entries $j\le n$ give the contribution of the first $n$
axial vector multiplets. (In the text, HW2(UV-fit) is not considered further because
it is a poor fit to IR data, and the SS model is discarded because of its wrong UV behavior.)
}
\label{tab:amuj}
\end{table}

\section{Conclusion}

In the present study, we have calculated the axial vector meson contributions arising from the Chern-Simons action
in the holographic QCD models, for which we had re-evaluated the pion-pole contribution to
the anomalous magnetic moment of the muon in Ref.~\cite{Leutgeb:2019zpq}.
We found that the infinite tower of axial vector mesons present in all these models
leads to a large-momentum behavior that in the case of hard-wall models
matches the MV-SDC \cite{Melnikov:2003xd} concurrently with the SDC for the pion TFF.

At low energies, we found that the holographic QCD models, now including the top-down SS model for 
low-energy large-$N_c$ QCD, reproduce well the experimentally determined shape of the $f_1(1285)$ TFF.
The SS and HW2 models also agree with its normalization, which is related to $\tilde\Gamma_{\gamma\gamma}$,
while the HW1 model overestimates the latter.
On the other hand, the HW2 model saturates the MV-SDC only at the level of 62\%,
whereas the HW1 model does so completely.

In Ref.~\cite{Leutgeb:2019zpq} we have found that the HW1 and HW2 models bracket the low-energy results
for the pion TFF, giving also the high and low ends of our results for $a_\mu^{\pi^0}=5.9(2)\times 10^{-10}$.
Evaluating the axial vector contributions we obtained $a_\mu^\mathrm{AV}=4.06\times 10^{-10}$ and $2.87\times 10^{-10}$,
respectively, 57 and 58\% of which are coming from the longitudinal contribution responsible for the MV-SDC. 
This is somewhat larger than the estimate of $\Delta a_\mu^\mathrm{LSDC}$ obtained recently in \cite{Colangelo:2019lpu,*Colangelo:2019uex}.

The HW1 result, which gives a similarly high result while completely saturating
the MV-SDC, appears to significantly overestimate the measured two-photon rate of
$f_1(1285)$, so that one might favor the smaller result coming from the HW2 model. 
On the other hand, real QCD has excited axial vector mesons that are lighter than
predicted by the holographic models (see Table \ref{avmasses}), and those will also
contribute to $a_\mu^\mathrm{AV}$. Moreover, away from the chiral limit excited pseudoscalar
mesons have to be included, as pointed out in \cite{Colangelo:2019lpu,*Colangelo:2019uex}.
This could mean that the final result for $\Delta a_\mu$ from the pseudoscalar and axial vector sector
might indeed be in between the HW1 and HW2 results.

We have also considered the possibility of using the holographic results for the axial vector TFF
as a phenomenological interpolator, where the normalization is fitted to the experimental results for $\gamma\gamma^*\to f_1$.
Consistency of the results for $f_1(1285)$ and $f_1(1420)$ leads to a mixing angle $\phi_f\approx 20^\circ$
away from ideal mixing, in agreement with other phenomenological studies.
Using the resulting overall normalization, we arrived at the result  
$a_\mu^\mathrm{AV1}=1.74(40)\times 10^{-10}$ for the lightest axial vector meson multiplet.
This is also significantly larger than was obtained in previous recent studies using various ans\"atze for the TFF 
\cite{Jegerlehner:2017gek,Pauk:2014rta}, which gave\footnote{%
Two very recent papers \cite{Roig:2019reh,Dorokhov:2019tjc} arrived at even lower results.}
$a_\mu^\mathrm{AV1}=(0.4$ - $1)\times 10^{-10}$.
Since our study suggests that $a_\mu^\mathrm{AV1}$ may account for only 80\% of the axial vector meson 
sector (perhaps even less, since the holographic models overestimate the mass of excited axial vector mesons),
also this more data-based approach suggests a contribution of (at least) $a_\mu^\mathrm{AV}=2.2(5)\times 10^{-10}$,
close to the pristine result of the HW2 model.

In summary, our holographic results underline
the numerical importance of axial vector contributions,
and their role in
satisfying the MV-SDC \cite{Melnikov:2003xd}. 
However, our results for the effect of the latter, which
can be attributed to 57-58\% of the axial vector contribution, are significantly smaller than
what is obtained by the simple MV model (when updated to modern input data), where 
in the pseudoscalar contributions one structure function is artificially kept fixed to its on-shell value.
More importantly, the holographic QCD calculation indicates that the MV model is not the correct way
to implement the MV-SDC, but that additional degrees of freedom are needed.
On the other hand, our results are larger than (albeit not too far above)
the estimate in \cite{Colangelo:2019lpu,*Colangelo:2019uex} from a model involving an infinite tower
of pseudoscalar excitations. Moreover, the axial vector TFF obtained in holographic QCD provide
a well motivated model for the double-virtual case which differs strongly from a simple dipole ansatz,
suggesting that previous estimates of the axial vector contribution to $a_\mu$ are significantly too small.

%{\em Note Added:} 
After completion of this work, Ref.~\cite{Cappiello:2019hwh} appeared, which also has worked out
the contribution of axial vector mesons in the HW2 model, in essence agreeing with our findings,
but employing different sets of parameters %(including what we called HW2(UV-fit), but
(with decay constants chosen differently for a partition
in $\pi^0/a_1$, $\eta/f_1$, and $\eta'/f_1'$ sectors\footnote{%As a consequence of their parameter choice, 
Treating $f_1$ and $f_1'$ in line with $\eta$ and $\eta'$, Ref.~\cite{Cappiello:2019hwh} found the $f_1'$ axial vector mesons
to give the largest contribution, whereas both ideal mixing and the mixing with $\phi_f=20.4^\circ$ we inferred from
experimental data should have the largest contribution coming from $f_1$, in particular given the higher
mass of $f_1'$. In our models, where the axial vector meson multiplets have strictly degenerate masses
as given in Table \ref{avmasses}, mixing with $\phi_f=20.4^\circ$ implies a partition of
$25\%,49\%,26\%$ for the $a_1,f_1,f_1'$ contributions, whereas in real QCD the $f_1'$ contribution should be somewhat reduced by its higher mass.}).
Their choice ``Set 1'' corresponds roughly to our treatment of the HW2 model, where a fit of $f_\pi$ and $m_\rho$
implies that only $62\%$ of the MV SDC is reached; ``Set 2'' corresponds to what we called HW2(UV-fit), where
100\% of the MV SDC is satisfied at the expense of a much too large $\rho$ meson mass, whereas we have employed
the HW1 model to be able to match SDCs as well as the low-energy parameters.
Since Ref.~\cite{Cappiello:2019hwh} included the pseudoscalar contributions in their final results,
let us point out that our results are to be added to our previous results of the pseudoscalar pole
contribution calculated before in Ref.~\cite{Leutgeb:2019zpq}. To facilitate a comparison of
our results with those of Ref.~\cite{Cappiello:2019hwh}, our final results for the pseudoscalar
plus axial vector sector are rendered in Table \ref{tab:amutotal}, where the last column corresponds to the adjustment of the axial vector contribution to match L3 data.
As stated above, this downscaled result could be viewed as 
an extrapolation to real QCD, where in contrast to the chiral large-$N_c$ limit
excited pseudoscalars are also contributing and which one would then have to add in \cite{Colangelo:2019lpu,*Colangelo:2019uex}.

\begin{table}[h]
\bigskip
\begin{tabular}{l|c|c|c}
\toprule
 & HW1 & HW2 & Extrapolation\\
\colrule
% $a_\mu^\mathrm{PS}[\pi^0+\eta+\eta']\times 10^{10}$ & \st{9.22 [6.13+1.67+1.42]} & 8.37 \st{[5.92+1.59+1.34]} & \st{8.8(4)} \\
$a_\mu^\mathrm{PS}[\pi^0+\eta+\eta']\times 10^{10}$ & {9.90 [6.52+1.82+1.56]} & 8.37 {[5.66+1.48+1.24]} & {9.1(8)} \\
$a_\mu^\mathrm{AV}[L+T]\times 10^{10}$ & 4.06 [2.32+1.74] & 2.87 [1.66+1.20] & 2.2(5) [1.3(3)+0.9(2)] \\
\colrule
% $a_\mu^\mathrm{PS+AV}\times 10^{10}$ & \st{13.3} & 11.2  & \st{11.0(6)}\\
$a_\mu^\mathrm{PS+AV}\times 10^{10}$ & {14.0} & 11.2  & 11.3(1.3)\\
\botrule
\end{tabular}
\caption{Summary of the results of our previous calculation of the pseudoscalar pole contribution of Ref.~\cite{Leutgeb:2019zpq}
and the results obtained here for the contribution of the infinite tower of axial vector mesons in the holographic models HW1 and HW2,
where the last column (Extrapolation) contains the span of the (ground-state)
pseudoscalar pole contributions in the two models, and the rescaled result for the axial
vector contributions to match L3 data for $f_1(1285)$ and $f_1(1420)$. 
}
\label{tab:amutotal}
\end{table}

\begin{acknowledgments}
We are indebted to Massimiliano Procura and Jan L\"udtke for most useful discussions 
and suggestions, and for
help with the results of Refs.~\cite{Colangelo:2015ama,Colangelo:2017fiz}.
We cordially thank the authors of Ref.~\cite{Cappiello:2019hwh} for correspondence
after the publication of our respective works,
and also Hans Bijnens, Gilberto Colangelo, Franziska Hagelstein, and Martin Hoferichter for comments on the manuscript.
J.~L.\ was supported by the FWF doctoral program
Particles \& Interactions, project no. W1252-N27.
\end{acknowledgments}

%\newpage
\appendix
\section{Helicity amplitudes for $\gamma^*\gamma^*\to\mathcal{A}$}\label{AppMhel}

For future potential tests of the holographic predictions in the double-virtual case, 
we list here the helicity
amplitudes for photoproduction of axial vector mesons,
generalizing the formulae given in Eq.~(C21) of Ref.~\cite{Pascalutsa:2012pr} from
a symmetric structure function $A(Q_1^2,Q_2^2)$ to the asymmetric one
appearing in (\ref{AHW}). Note that the generic form of the amplitude admits
one further structure function (denoted by $C(Q_1^2,Q_2^2)$ in Ref.~\cite{Roig:2019reh}),
which vanishes in the holographic result (\ref{calMa}).

With the definitions \cite{Pascalutsa:2012pr}
\bea
&&\nu:=q_{(1)}\cdot q_{(2)}=\frac12(M_A^2+Q_1^2+Q_2^2),\nonumber\\
&&X:=\nu^2-Q_1^2 Q_2^2,
\eea
and the overall constant $C:=\mathrm{tr}(\mathcal{Q}^2 t^a){N_c}/({4\pi^2})$,
the holographic result for the amplitude (\ref{calMa}) in terms of the functions
$A\equiv A(Q_1^2,Q_2^2)$ and $\bar A\equiv A(Q_2^2,Q_1^2)$ contains the following
nonzero $\gamma^*\gamma^*\to\mathcal{A}$ helicity amplitudes:
\bea
-i\mathcal{M}_{++}/C&=& \frac{\nu}{M_A} (Q_1^2 A-Q_2^2 \bar A)- \frac{Q_1^2 Q_2^2}{M_A} (A-\bar A), \\
-i\mathcal{M}_{0+}/C&=& Q_1( \nu A+ Q_2^2 \bar A),\\
-i\mathcal{M}_{-0}/C&=& Q_2( \nu \bar A+Q_1^2 A),
\eea
where the first two indices refer to the helicities of the two virtual photons.

The structure functions $F^{(0)}_{\mathcal{A}\gamma^*\gamma^*}(Q_1^2,Q_2^2)$
and $F^{(1)}_{\mathcal{A}\gamma^*\gamma^*}(Q_1^2,Q_2^2)$ defined in Eq.~(C14) of Ref.~\cite{Pascalutsa:2012pr}
are proportional to $\mathcal{M}_{++}/[(Q_1^2-Q_2^2)\nu/M_A^3]$ and
$\mathcal{M}_{0+}/[Q_1X/(\nu M_A^2)]$, respectively.

\section{Integral representation of $a_\mu^\mathrm{AV}$}\label{AppIntegrals}

With the method of Gegenbauer polynomials described in Ref.~\cite{Jegerlehner:2009ry}
we have obtained
\begin{align}
a_{\mu}^\mathrm{AV}= & -\frac{2\alpha^{3}}{3\pi^{2}}\int_{0}^{\infty}dQ_{1}dQ_{2}\int_{-1}^{+1}d\tau\sqrt{1-\tau^{2}}Q_{1}^{3}Q_{2}^{3}\left(K_{1}+K_{2}\right),
\end{align}
with $K_1$ the integral kernel for the $s$-channel reading 
{\footnotesize
\begin{align}
K_{1}= & \frac{A(Q_{3}^{2},0)\left(Q_{1}^{2}A(Q_{1}^{2},Q_{2}^{2})+Q_{2}^{2}A(Q_{2}^{2},Q_{1}^{2})\right)}{2Q_{1}Q_{2}Q_{3}^{2}m_{\mu}^{2}M_{A}^{2}}\left[\tau\left(Q_{2}^{2}(4\sigma_{2}^{E}+\left(\sigma_{2}^{E}\right)^{2}-5)-8m_{\mu}^{2}\right)\right.\nonumber \\
 & \left.-4Q_{2}Q_{1}\left(-4\left(\tau^{2}-1\right)Xm_{\mu}^{2}+2Q_{2}^{2}X-\sigma_{1}^{E}-\sigma_{2}^{E}+2\right)-8Q_{2}Q_{1}^{3}X+Q_{1}^{2}\tau\left(-16Q_{2}^{2}X+4\sigma_{1}^{E}+\left(\sigma_{1}^{E}\right)^{2}-5\right)\right]\nonumber \\
+&\frac{A(Q_{1}^{2},Q_{2}^{2})A(Q_{3}^{2},0)}{2Q_{1}Q_{2}^{2}Q_{3}^{2}m_{\mu}^{2}\left(M_{A}^{2}+Q_{3}^{2}\right)}\left[Q_{2}^{3}Q_{1}^{2}\tau\left(-8Q_{2}^{2}X+2\sigma_{1}^{E}+\left(\sigma_{1}^{E}\right)^{2}-2\sigma_{2}^{E}\tau^{2}+8\sigma_{2}^{E}+\left(\sigma_{2}^{E}\right)^{2}\tau^{2}+\left(\sigma_{2}^{E}\right)^{2}+\tau^{2}-12\right)\right.\nonumber \\
 & \left.-4m_{\mu}^{2}\left(Q_{2}^{3}\tau+Q_{1}^{3}\left(1-4Q_{2}^{2}\tau^{2}X\right)+Q_{2}^{2}Q_{1}\left(4Q_{2}^{2}X+\tau^{2}\right)-4Q_{2}Q_{1}^{4}\tau X+Q_{2}Q_{1}^{2}\tau\left(4Q_{2}^{2}X+3\right)\right)\right.\nonumber \\
 & \left.+2Q_{2}^{4}Q_{1}\left(-2Q_{2}^{2}X+\sigma_{1}^{E}-\sigma_{2}^{E}\tau^{2}+2\sigma_{2}^{E}+\left(\sigma_{2}^{E}\right)^{2}\tau^{2}-3\right)+2Q_{2}^{2}Q_{1}^{3}\left(-6Q_{2}^{2}X+\sigma_{1}^{E}+2\sigma_{2}^{E}-3\right)\right.\nonumber \\
 & \left.-4Q_{2}Q_{1}^{4}\tau\left(4Q_{2}^{2}X-\sigma_{1}^{E}+1\right)+Q_{1}^{5}\left(-8Q_{2}^{2}X+2\sigma_{1}^{E}-2\right)+Q_{2}^{5}(\left(\sigma_{2}^{E}\right)^{2}-1)\tau\right]\nonumber \\
+&\frac{A(Q_{2}^{2},Q_{1}^{2})A(Q_{3}^{2},0)}{2Q_{1}^{2}Q_{2}Q_{3}^{2}m_{\mu}^{2}\left(M_{A}^{2}+Q_{3}^{2}\right)}\left[2Q_{1}^{2}\left(-2Q_{2}\tau^{2}m_{\mu}^{2}+Q_{2}^{3}\left(8\tau^{2}Xm_{\mu}^{2}+2\sigma_{1}^{E}+\sigma_{2}^{E}-3\right)-4Q_{2}^{5}X\right)-4Q_{2}Q_{1}^{6}X   \phantom{\biggr]}\right.\nonumber \\
 & \left.-2Q_{2}Q_{1}^{4}\left(8Xm_{\mu}^{2}+6Q_{2}^{2}X+\sigma_{1}^{E}\tau^{2}-2\sigma_{1}^{E}-\left(\sigma_{1}^{E}\right)^{2}\tau^{2}-\sigma_{2}^{E}+3\right)\right.\nonumber \\
 & +\left.Q_{1}^{3}\tau\left(Q_{2}^{2}\left(-16Q_{2}^{2}X-2\sigma_{1}^{E}\left(\tau^{2}-4\right)+\left(\sigma_{1}^{E}\right)^{2}\left(\tau^{2}+1\right)+2\sigma_{2}^{E}+\left(\sigma_{2}^{E}\right)^{2}+\tau^{2}-12\right)-4m_{\mu}^{2}\left(4Q_{2}^{2}X+1\right)\right)\right.\nonumber \\
 & \left.-2Q_{2}^{3}\left(2m_{\mu}^{2}-Q_{2}^{2}(\sigma_{2}^{E}-1)\right)+4Q_{2}^{2}Q_{1}\tau\left(m_{\mu}^{2}\left(4Q_{2}^{2}X-3\right)+Q_{2}^{2}(\sigma_{2}^{E}-1)\right)-Q_{1}^{5}\tau\left(8Q_{2}^{2}X-\left(\sigma_{1}^{E}\right)^{2}+1\right)\right],
\end{align}
}
and $K_2$ for the $t$- and $u$-channels,
{\footnotesize
\begin{align}
K_{2}= & \frac{A(Q_{2}^{2},0)\left(Q_{3}^{2}A(Q_{3}^{2},Q_{1}^{2})+Q_{1}^{2}A(Q_{1}^{2},Q_{3}^{2})\right)}{Q_{1}Q_{2}Q_{3}^{2}m_{\mu}^{2}M_{A}^{2}}\times\nonumber \\
 & \left.\left(-4\tau m_{\mu}^{2}-4Q_{1}Q_{2}\left(\tau^{2}-1\right)\left(-4Xm_{\mu}^{2}+2Q_{2}^{2}X-\sigma_{1}^{E}+1\right)+Q_{1}^{2}(\left(\sigma_{1}^{E}\right)^{2}-1)\tau\right)\right.\nonumber \\
+&\frac{A(Q_{1}^{2},Q_{3}^{2})A(Q_{2}^{2},0)}{Q_{1}Q_{2}Q_{3}^{2}m_{\mu}^{2}\left(M_{A}^{2}+Q_{2}^{2}\right)}\left[2Q_{2}Q_{1}^{3}\left(4Q_{2}^{2}\left(\tau^{2}+1\right)X-4\sigma_{1}^{E}\tau^{2}+\left(\sigma_{1}^{E}\right)^{2}\tau^{2}-\sigma_{2}^{E}+3\tau^{2}+1\right)\right.\nonumber \\
 & \left.+Q_{2}^{2}Q_{1}^{2}\tau(-2\sigma_{1}^{E}+\left(\sigma_{1}^{E}\right)^{2}-6\sigma_{2}^{E}-\left(\sigma_{2}^{E}\right)^{2}+8)+Q_{1}^{4}\tau\left(16Q_{2}^{2}X-6\sigma_{1}^{E}+\left(\sigma_{1}^{E}\right)^{2}+5\right)\right.\nonumber \\
 & \left.-4m_{\mu}^{2}\left(Q_{2}^{2}\tau+4Q_{2}Q_{1}^{3}\left(\tau^{2}+1\right)X+2Q_{1}^{2}\tau\left(6Q_{2}^{2}X-1\right)+Q_{2}Q_{1}\left(4Q_{2}^{2}X-1\right)\right)\right.\nonumber \\
 & \left.+2Q_{2}^{3}Q_{1}\left(-2Q_{2}^{2}X+\sigma_{1}^{E}-3\sigma_{2}^{E}+2\right)+Q_{2}^{4}(\left(\sigma_{2}^{E}\right)^{2}-1)\tau+4Q_{2}Q_{1}^{5}X\right]\nonumber \\
-&\frac{A(Q_{2}^{2},0)A(Q_{3}^{2},Q_{1}^{2})}{Q_{1}^{2}Q_{2}Q_{3}^{2}m_{\mu}^{2}\left(M_{A}^{2}+Q_{2}^{2}\right)}\left[10Q_{2}^{4}Q_{1}(\sigma_{2}^{E}-1)\tau+2Q_{2}^{5}(\sigma_{2}^{E}-1)-4Q_{2}Q_{1}^{6}X   \phantom{\biggr]}\right.\nonumber \\
 & \left.+4m_{\mu}^{2}\left\{4Q_{2}Q_{1}^{4}\left(\tau^{2}+1\right)X+2Q_{1}^{3}\tau\left(2Q_{2}^{2}\left(2\tau^{2}+3\right)X-1\right)\right.\right.\nonumber \\
 &\qquad \left.\left.+Q_{2}Q_{1}^{2}\left(4Q_{2}^{2}\left(3\tau^{2}+1\right)X-4\tau^{2}-1\right)+2Q_{2}^{2}Q_{1}\tau\left(2Q_{2}^{2}X-1\right)-Q_{2}^{3}\right\}\right.\nonumber \\
 & \left.-Q_{2}^{2}Q_{1}^{3}\tau\left(16Q_{2}^{2}\left(\tau^{2}+1\right)X-2\sigma_{1}^{E}\left(5\tau^{2}+2\right)+\left(\sigma_{1}^{E}\right)^{2}\left(\tau^{2}+1\right)-10\sigma_{2}^{E}-\left(\sigma_{2}^{E}\right)^{2}+9\tau^{2}+14\right)\right.\nonumber \\
 & \left.-Q_{1}^{5}\tau\left(24Q_{2}^{2}X-6\sigma_{1}^{E}+\left(\sigma_{1}^{E}\right)^{2}+5\right)+2Q_{2}^{3}Q_{1}^{2}\left(\tau^{2}\left(-4Q_{2}^{2}X+2\sigma_{1}^{E}-9\right)+\sigma_{2}^{E}\left(6\tau^{2}+2\right)+\left(\sigma_{2}^{E}\right)^{2}\tau^{2}-2\right)\right.\nonumber \\
 & \left.-2Q_{2}Q_{1}^{4}\left(2Q_{2}^{2}\left(10\tau^{2}+1\right)X-8\sigma_{1}^{E}\tau^{2}+\left(\sigma_{1}^{E}\right)^{2}\tau^{2}-\sigma_{2}^{E}+7\tau^{2}+1\right)\right],
\end{align}
}

\noindent for one axial vector meson multiplet with mass $M_A$ and form factors as defined in (\ref{AHW}),
but with the prefactor $\frac{N_c}{4\pi^2}\mathrm{tr}(\mathcal{Q}^2 t^a)$ of (\ref{calMa}) included in $A$.
In the integral kernels we have used the notation of \cite{Colangelo:2015ama,Colangelo:2017fiz},
\begin{align}
Q_{3}^{2}= & Q_{1}^{2}+2Q_{1}Q_{2}\tau+Q_{2}^{2}, \quad X=  \frac{1}{Q_{1}Q_{2}x}\arctan\left(\frac{zx}{1-z\tau}\right),\quad x=\sqrt{1-\tau^{2}},\nonumber\\
z= & \frac{Q_{1}Q_{2}}{4m_{\mu}^{2}}\left(1-\sigma_{1}^{E}\right)\left(1-\sigma_{2}^{E}\right),\quad\sigma_{i}^{E}=\sqrt{1+\frac{4m_{\mu}^{2}}{Q_{i}^{2}}}.
\end{align}

The sum over the infinite tower of axial vector mesons corresponds to replacing the products of two form factors $A$ and the denominator of the axial vector meson propagator
by one double-integral expression involving the bulk-to-bulk propagator as discussed in Sect.~\ref{sec:SDC}, e.g.,
\bea
&&\sum_{n=1}^\infty \frac{A_{(n)}(Q_1^2,Q_2^2)A_{(n)}(Q_3^2,0)}{Q_3^2+(M^A_n)^2}\nonumber\\
&&=\frac{4g_5^2}{Q_1^2 Q_3^2}\int_0^{z_0} dz 
\left[ \frac{d}{dz} \mathcal{J}(Q_1,z) \right]\mathcal{J}(Q_2,z)
\int_0^{z_0} dz' 
\left[ \frac{d}{dz'} \mathcal{J}(Q_3,z') \right]
G^A(Q_3;z,z'),
\eea
which in longitudinal contributions reduces to
\bea
&&\sum_{n=1}^\infty \frac{A_{(n)}(Q_1^2,Q_2^2)A_{(n)}(Q_3^2,0)}{(M^A_n)^2}\nonumber\\
&&=\frac{4g_5^2}{Q_1^2 Q_3^2}\int_0^{z_0} dz 
\left[ \frac{d}{dz} \mathcal{J}(Q_1,z) \right]\mathcal{J}(Q_2,z)
\int_0^{z_0} dz' 
\left[ \frac{d}{dz'} \mathcal{J}(Q_3,z') \right]
G^A(0;z,z').
\eea

As mentioned above, we have checked that our results agree upon integration with those obtained in an alternative derivation using
the formalism of
Refs.~\cite{Colangelo:2015ama,Colangelo:2017fiz}. 
The latter agree with those given in Appendix C of Ref.~\cite{Cappiello:2019hwh}. %up to typos

\raggedright
\bibliographystyle{JHEP}
\bibliography{hlbl}

\end{document}